\documentclass[aps,prl,twocolumn,showpacs,superscriptaddress]{revtex4}
\usepackage{amsmath}
\usepackage{latexsym}
\usepackage{amssymb}
\usepackage{graphicx}
\usepackage{bm}

\newcommand{\pdag}{{\phantom{\dagger}}}
\newcommand{\bq}{\begin{equation}}
\newcommand{\eq}{\end{equation}}
\newcommand{\bn}{\begin{eqnarray}}
\newcommand{\en}{\end{eqnarray}}

\begin{document}

\title{Qubit measurement by a quantum point contact: a quantum
Langevin equation approach}

\author{Bing Dong}
\affiliation{Department of Physics, Shanghai Jiaotong University, 1954 Huashan Road,
Shanghai 200030, China}
\affiliation{Department of Physics and Engineering Physics, Stevens Institute of
Technology, Hoboken, New Jersey 07030, USA}

\author{Norman J. M. Horing}
\affiliation{Department of Physics and Engineering Physics, Stevens Institute of
Technology, Hoboken, New Jersey 07030, USA}

\author{X. L. Lei}
\affiliation{Department of Physics, Shanghai Jiaotong University, 1954 Huashan Road,
Shanghai 200030, China}

\begin{abstract}

We employ a quantum Langevin equation approach to establish Bloch-type dynamical equations, 
on a fully microscopic basis, to investigate the measurement of
the state of a coupled quantum dot qubit by a nearby quantum point contact. The ensuing Bloch 
equations allow us to analyze qubit relaxation and decoherence induced by measurement,
and also determine the noise spectrum of the meter output current with the help of a quantum 
regression
theorem, at arbitrary bias-voltage and temperature. Our analyses provide a clear resolution 
of a recent debate concerning the occurrence of a quantum oscillation peak in the noise 
spectrum.

\end{abstract}


\today

\maketitle

The quantum point contact (QPC) has been proposed and designed as an
efficient, practical weak continuous measurement device 
(meter)\cite{Buks,Schoelkopf,Nakamura,Mozyrsky,Buttiker,Korotkov,Goan}. Recently, theoretical 
analyses of the detector noise spectrum have been
carried out under arbitrary bias-voltage and temperature conditions using a perturbative
Green's function method\cite{Shnirman,Bulaevskii}, also a quantum jump technique\cite{Stace},
and a Gurvitz-type master equation approach with a multi-subspace ansatz\cite{Li}, in which
some results have led to disagreements and remain inconsistent\cite{dispute}. Therefore, this 
problem requires further study.

Here, we revisit this issue employing a recently developed generic quantum
Langevin equation approach\cite{Dong}, which we approximate to establish a set of Bloch-type 
dynamical equations describing the time evolutions of qubit variables explicitly in terms of 
the response and correlation functions of the detector variables. These Bloch
equations provide analytical expressions for the bias-voltage and
temperature $T$-dependent
relaxation and decoherence of the qubit induced by measurement. We then evaluate the
detector current and frequency-independent shot noise within 2nd-order perturbation theory,
and they are shown to fulfill the nonequilibrium fluctuation-dissipation (NFD) 
theorem\cite{NFDT}.
Finally, we calculate the frequency-dependent noise using a quantum regression theorem (QRT) 
based on the derived Bloch equations\cite{QRT}.

The system we examine is a double quantum dot charged qubit $c_{1}$ and $c_{2}$ coupled with 
a nearby low-transparency QPC detector. In
order to properly account for dissipative effects in the qubit evolution, we
write the total Hamiltonian of the system in terms of the eigenstate basis of the
qubit, i.e. $c_{\alpha }=\cos \frac{\vartheta }{2}c_{1}+\sin \frac{\vartheta }{2}c_{2}$,
$c_{\beta }=\sin \frac{\vartheta }{2}c_{1}-\cos \frac{\vartheta }{2}c_{2}$, with $\vartheta
=\tan ^{-1}\frac{\Omega }{\delta }$ ($\Omega $ is the tunnel-coupling between two
quantum dots with energies $\pm \delta $):
\begin{eqnarray}
H_{q} &=&\Delta \sigma _{z},\,\,\,H_{\mathrm{m}}=\sum_{\eta ,\mathbf{k}}\varepsilon _{\eta 
\mathbf{k}}c_{\eta \mathbf{k}}^{\dag }c_{\eta \mathbf{k}}^{\pdag},  \label{H} \\
H_{\mathrm{I}}
&=& \sum_{i=1,2}Q_{i}^{z}F_{Q^{z}}+Q_{i}^{+}F_{Q^{+}}+Q_{i}^{-}F_{Q^{-}}+Q_{i}^{\hat{1}} 
F_{Q^{\hat{1}}},
\nonumber
\end{eqnarray}
where $\Delta =2\sqrt{\delta ^{2}+\Omega ^{2}}$ measures the energy difference
between two eigen-energies, i.e. the Rabi frequency; 
$c_{\eta \mathbf{k}}^{\dagger }$ ($c_{\eta \mathbf{k}}$) is the electron creation 
(annihilation) operator in lead $\eta$($=L,R$) of the QPC with
a flat density of states $\rho _{\eta }$; $Q_{i}^{z}=-Q_{i}(t)\cos \vartheta $, 
$Q_{i}^{+}=Q_{i}^{-}=-\frac{1}{2}Q_{i}(t)\sin \vartheta $, $Q_{i}^{\hat{1}}=\frac{W}{\chi 
}Q_{i}(t)$ with $Q_{1}(t)\equiv \chi \sum_{\mathbf{k},\mathbf{q}}c_{L\mathbf{k}}^{\dag 
}c_{R\mathbf{q}}^{\pdag}$ and $Q_{2}(t)\equiv
Q_{1}^{\dagger }(t)$ are the generalized co-ordinates; $W$ and $\chi $ are the direct and the 
qubit-modified tunneling matrix elements of the QPC,
respectively, which are taken as constants; $F_{Q^{z}}=\hat\sigma _{z}=\frac{1}{2}(c_{\alpha
}^{\dagger }c_{\alpha }^{\pdag}-c_{\beta }^{\dagger }c_{\beta }^{\pdag})$, $F_{Q^{\mp 
}}=\hat\sigma _{\pm} = c_{\alpha (\beta )}^{\dagger }c_{\beta (\alpha )}^{\pdag}$, and 
$F_{Q^{\hat{1}}}=\hat{1}$ (the unit operator) are the generalized forces.
Here, the terms $Q^{\pm }F_{Q^{\pm }}$ describe \emph{energy-exchange} (\emph{inelastic})
processes during measurement, in which a qubit-state transition takes place from one
eigenstate to the other one due to the measurement. Note that our setup, Eq.~(\ref{H}), is 
similar with those of previous studies\cite{Korotkov,Goan,Shnirman,Stace}.

In our derivation of Bloch-type dynamical equations for the qubit
variables, operators of the qubit and the detector are first expressed formally by 
integration of
their Heisenberg equations of motion (EOM), exactly to all orders in the tunnel coupling and
qubit-detector coupling constants, $W$ and $\chi$. Next, under the assumption that the time
scale of
decay processes is much slower than that of free evolutions, we replace the time-dependent
operators involved in the integrals of these EOM's approximately in terms of their free
evolutions (Markov approximation). Finally, these EOM's are expanded in powers of coupling 
constants up to second
order, taking the convenient and compact form\cite{Dong}:
\begin{widetext}
\begin{subequations}
\label{eom1}
\bn
\langle \dot {\hat \sigma}_z(t) \rangle &=& -{1\over 2} \sin \vartheta \int_{-\infty}^t d\tau 
\sum_{c,n,l} \Bigl \{ \frac{1}{2} \theta(\tau) \langle [Q_{no}(t), Q_{lo}^{c}(t')]_{+} 
\rangle_{m} \langle [\hat\sigma_{+}^o(t) - \hat\sigma_{-}^o(t), F_{Q^c}^o(t')]_{-} 
\rangle_{q} \cr
&& +{1\over 2} \theta(\tau) \langle [Q_{no}(t), Q_{lo}^{c}(t')]_{-} \rangle_{m} \langle 
[\hat\sigma_{+}^o(t) - \hat\sigma_{-}^o(t), F_{Q^c}^o(t')]_{+} \rangle_{q} \Bigr \}, 
\label{eom:sz1}
\en
\bn
\langle \dot {\hat\sigma}_{\pm}(t) \rangle &=& \pm i \Delta \langle \hat\sigma_\pm(t) \rangle 
\mp \int_{-\infty}^t d\tau \sum_{c,n,l} \Bigl \{ \frac{1}{2} \theta(\tau) \langle [Q_{no}(t), 
Q_{lo}^{c}(t')]_{+} \rangle_{m} \langle [\cos \vartheta \hat\sigma_{\pm}^o(t)- \sin \vartheta 
\hat\sigma_z^o(t), F_{Q^c}^o(t')]_{-} \rangle_{q} \cr
&& + \frac{1}{2} \theta(\tau) \langle [Q_{no}(t), Q_{lo}^{c}(t')]_{-} \rangle_{m} \langle 
[\cos \vartheta \hat\sigma_{\pm}^o(t)- \sin \vartheta \hat\sigma_z^o(t), F_{Q^c}^o(t')]_{+} 
\rangle_{q} \Bigr \},
\label{eom:s+1}
\en
\end{subequations}
\end{widetext}
where the summation notation $c$ is over $\{z,\pm ,{\hat 1}\}$, and $n,l$ are over $\{1,2\}$; 
the
statistical average $\langle \cdots \rangle _{m(q)}$ is in regard to the detector (qubit) 
variables; and
the super(sub)script \textquotedblleft $o$" signifies free evolution of the corresponding
variables; $\tau =t-t^{\prime }$. In conjunction with free evolution, the qubit dynamics are 
modified by the measurement process in
a way that relates to the response function, $R_{nl}(t,t^{\prime })$, and
correlation function, $C_{nl}(t,t^{\prime })$, of free meter variables, which are
defined as:
\begin{eqnarray}
R_{nl}(t,t^{\prime }) &=&{\frac{1}{2}}\theta (\tau )\langle \lbrack
Q_{no}(t),Q_{lo}(t^{\prime })]_{-}\rangle _{m},  \label{responsef} \\
C_{nl}(t,t^{\prime }) &=&{\frac{1}{2}}\theta (\tau )\langle \lbrack
Q_{no}(t),Q_{lo}(t^{\prime })]_{+}\rangle _{m}.  \label{correlationf}
\end{eqnarray}
The nonvanishing correlation functions can be readily expressed in terms of
reservoir Fermion distribution functions of the meter, and their Fourier transforms 
are\cite{Dong}:
\begin{equation}
R_{12/21}(\omega )=g_{1}(\omega \pm V),\,\,\,C_{12/21}(\omega
)=g_{1}T\varphi \left( \frac{V\pm \omega }{T}\right) , \label{RC}
\end{equation}
with $g_{1}=\pi \rho _{L}\rho _{R}\chi ^{2}/2$ and $\varphi (x)=x\coth (x/2)$. $V$ is the 
bias-voltage applied between the left and right leads of the meter. We use units with $\hbar 
=k_{B}=e=1$.

For long time scale of interest, making the replacement $\int_{-\infty }^{t}d\tau
\Longrightarrow \int_{-\infty }^{\infty }d\tau $, these EOM's, Eq.~(\ref{eom1}), can be 
further simplified as [here, we write $\langle \hat\sigma_{z(\pm)}\rangle\rightarrow 
\sigma_{z(\pm)}$]:
\begin{subequations}
\label{eom2}
\begin{eqnarray}
\dot{\sigma}_{z} &=&-\frac{1}{T_{1}}\sigma _{z}+\Gamma _{z}\sigma _{p}+c_{z},
\label{eom:sz2} \\
\dot{\sigma}_{\pm } &=&\pm i\Delta \sigma _{\pm }-\frac{1}{T_{2}}\sigma_{\pm }+\Gamma 
_{d}\sigma _{\mp }+\Gamma _{\parallel }\sigma_{z}+c_{\parallel },  \label{eom:s+2}
\end{eqnarray}
\end{subequations}
with $\sigma _{p(m)}=\sigma _{+}\pm \sigma _{-}$, $\Gamma _{z}=\frac{1}{4}
\sin 2\vartheta C^{+}(0)$, $\Gamma _{d}=\frac{1}{2T_{1}}$, $\Gamma _{\parallel} = 
\frac{1}{2}\sin
2\vartheta C^{+}(\Delta )$, $c_{z}=-\frac{1}{2}\sin ^{2}\vartheta R^{+}(\Delta )$, 
$c_{\parallel }=\frac{1}{4}\sin 2\vartheta R^{+}(\Delta )$ [$C(R)^{\pm }=C(R)_{12}\pm 
C(R)_{21}$], and the relaxation
rate $\frac{1}{T_{1}}$ and decoherence rate $\frac{1}{T_{2}}$ are given by:
\begin{equation}
\frac{1}{T_{1}}=\sin ^{2}\vartheta C^{+}(\Delta ),\,\,\,\,\frac{1}{T_{2}}=
\frac{1}{2T_{1}}+\cos ^{2}\vartheta C^{+}(0). \label{dissipation}
\end{equation}
It is readily seen that the relaxation time ($T_{1}$) stems
completely from \emph{inelastic} measurement events, which is conceptually consistent with 
the physical
definition of qubit relaxation. These \emph{inelastic} processes also contribute to qubit
decoherence with the partial rate, $1/2T_{1}$. In contrast, \emph{elastic} processes do not
induce relaxation but do contribute to pure decoherence with the partial rate 
$\cos^{2}\vartheta C^{+}(0)$. In the case of $\vartheta =0$ (no inter-dot hopping, 
$\Omega=0$), the
relaxation rate is, of course, zero, meaning that the qubit is completely localized, 
$\frac{1}{T_{2}}=C^{+}(0)$, and $\sigma _{z}^{\infty }=\pm \frac{1}{2}$ (depending on the 
initial state), $\sigma _{\pm }^{\infty }=0$; On the other hand, if $\vartheta =\pi /2$ [the 
symmetric (S) case, $\delta =0$],
$\frac{1}{T_{1}}=C^{+}(\Omega )$ and $\frac{1}{T_{2}}=\frac{1}{2T_{1}}$. We
have the general steady-state solutions of Eqs.~(\ref{eom:sz2}) and (\ref{eom:s+2}) as:
\begin{equation}
\sigma _{z}^{\infty }=-\frac{R^{+}(\Delta )}{2C^{+}(\Delta )},\,\,\,\sigma
_{\pm }^{\infty }=0.
\end{equation}
The Markov approximation employed in the derivation of Eqs.~(\ref{eom2}) requires that (1) 
measurement-induced decay of the qubit $\Gamma_d \lesssim \Delta$; and (2) rapid internal 
decay of the two electrodes, $1/\tau_c \gtrsim V$ (coarse-graining assumption), which implies 
that the noise spectrum in the following calculation is meaningful only for low frequencies 
$\omega \lesssim V$.    

The tunneling current operator through the meter is defined as the time rate
of change of charge density, $N_{L}=\sum_{\mathbf{k}}c_{L\mathbf{k}}^{\dagger
}c_{L\mathbf{k}},$ in the left lead:
\bn 
J_{L}(t) = \dot N_L &=& -i [(Q_1^z-Q_2^z) \hat\sigma_z +
(Q_1^+-Q_2^+) \hat\sigma_- \cr && +(Q_1^- - Q_2^-) \hat\sigma_+ +
(Q_1^{\hat 1}-Q_2^{\hat 1})\cdot \hat 1]. \label{tc} 
\en
Using linear-response theory we have 
\bn
I &=& \langle J_L(t) \rangle =-i\int_{-\infty}^t dt' \langle [ J_L(t), H_{\rm I}(t')]_- 
\rangle_{m,q} \cr
&=& 2A_1 R^-(0) + \Bigl [\frac{1}{2} \sin^2 \vartheta - \frac{W}{\chi} \sin \vartheta 
\sigma_p \Bigr ] R^-(\Delta) \cr
&& + \Bigl [ \sin^2 \vartheta \sigma_z - \frac{1}{4} \sin 2\vartheta \sigma_p \Bigr ] 
C^-(\Delta), \label{current} 
\en
with $A_1=\frac{1}{4} \cos^2 \vartheta - \frac{2W}{\chi} \cos \vartheta \sigma_z 
-\frac{W}{2\chi} \sin \vartheta \sigma_p + (\frac{W}{\chi})^2$. In the case of $\vartheta=0$, 
the steady-state current is $I_{l(r)}=V (g_0+g_1 \pm 2\sqrt{g_0g_1})$ with $g_0=2\pi \rho_L 
\rho_R W^2$, where $I_{l(r)}$ means the QPC current when dot 2(1) of the qubit is occupied by 
electron; For the S case $\delta=0$, we have 
\bq
I=g_0 V+ g_1 V \left ( 1 - \frac{\Delta}{V} \frac{C^-(\Delta)}{C^+(\Delta)}\right ). 
\label{Scurrent} 
\eq 

To address the noise spectrum of a ``symmetric" detector, we employ its definition as the 
Fourier transform of the current-current correlation function, $S(\tau )$, which can be 
calculated using linear-response theory,
\begin{equation}
S(\omega )=\int_{-\infty }^{\infty }d\tau e^{i\omega \tau }\frac{1}{2}
\langle \lbrack \delta J_{L}(t),\delta J_{L}(t^{\prime })]_{+}\rangle _{m,q},
\label{def:sn}
\end{equation}
with $\delta J_{L}(t)=J_{L}(t)-\langle J_{L}(t)\rangle $. It is well-known
that the shot noise spectrum consists of the $\omega $-independent Schottky
noise $S_{0}$ (the pedestal) and the $\omega $-dependent part. Substituting
the current operator of Eq.~(\ref{tc}) into Eq.~(\ref{def:sn}) and using the definitions 
Eqs.~(\ref{responsef})-(\ref{RC}), we obtain the $\omega
$-independent noise, $S_{0}$, as: 
\bn
S_0 &=& 2A_1^\infty C^+(0) + \Bigl [\frac{1}{2} \sin^2 \vartheta - \frac{W}{\chi} \sin 
\vartheta \sigma_p^\infty \Bigr ] C^+(\Delta) \cr
&& + \Bigl [ \sin^2 \vartheta \sigma_z^\infty - \frac{1}{4} \sin 2\vartheta \sigma_p^\infty 
\Bigr ] R^+(\Delta), \label{s0} 
\en
where the first terms on the right hand side of Eq.~(\ref{current}) at $t\rightarrow \infty $ 
and Eq.~(\ref{s0}) obey the NFD relation, while the other two terms
represent the generalized NFD relation due to energy-exchange processes involved in the 
course of measurement\cite{NFDT}.

From Eq.~(\ref{current}), it is clear that the time evolution of the qubit
variables determines the temporal behavior of the meter current, and is consequently 
responsible for
the frequency-relevant part of the noise spectrum. To calculate the $\omega$-dependent noise, 
it is thus necessary to evaluate the two-time correlation functions
$\sigma _{ab}(\tau )=\langle \sigma _{a}(t+\tau )\sigma _{b}(t)\rangle $ ($a,b=z,p,m$) using 
the QRT starting with the dynamic equations (\ref{eom2}). The QRT states that given
closed-form equations of motion of one-time averages of a set of operators $\mathcal{O}_{j}$:
\begin{equation}
\frac{d}{d\tau }\langle \mathcal{O}_{j}(t+\tau )\rangle
=\sum_{k}L_{j,k}\langle \mathcal{O}_{k}(t+\tau )\rangle +\lambda _{j},
\end{equation}
then the two-time averages of $\mathcal{O}_{j}$ with any other operator, $\mathcal{P}$, also 
obey the same equations\cite{QRT}:
\begin{equation}
\frac{d}{d\tau }\langle \mathcal{O}_{j}(t+\tau )\mathcal{P}(t)\rangle
=\sum_{k}L_{j,k}\langle \mathcal{O}_{k}(t+\tau )\mathcal{P}(t)\rangle
+\lambda _{j}\langle \mathcal{P}(t)\rangle .  \label{QRT}
\end{equation}
Because the solutions of interest here involve long time behavior after the
initial turn-on, we replace the \emph{initial values} of $\sigma _{ab}(0)=\langle
\sigma _{a}(t)\sigma _{b}(t)\rangle$ and the last term $\left\langle 
\mathcal{P}(t)\right\rangle$ in Eq.~(\ref{QRT}) with their respective \emph{steady-state 
values}. Once these equations for $\sigma _{ab}(\tau )$
are solved and substituted into the definition, Eq.(\ref{def:sn}), we
calculate the frequency-dependent noise spectrum $S(\omega)$.

There is no general analytic expression for the noise spectrum, except in
two special cases: In the absence of coupling, $\Omega =0$, 
$I(t)=[\frac{1}{2}+2(\frac{W}{\chi})^{2}]R^{-}(0)-\frac{4W}{\chi }\sigma _{z}(t)$, but 
$\dot{\sigma}_{z}=0$, therefore the total noise is independent of $\omega$ with 
$S=I_{l(r)}\coth (V/2T)$; On the contrary, for the S system $\delta=0$, we have 
$I(t)=I_{0}+I_{1}(t)=(g_{0}+g_{1})V-A\sigma_{p}(t) + B\sigma _{z}(t)$, with $A=\frac{4W}{\chi 
}g_{1}V$ and $B=C^{-}(\Delta )$. Moreover, since $\sigma _{z}(t)$ is decoupled from 
$\sigma_{\pm }(t)$ in Eqs.~(\ref{eom2}) due to $\Gamma_{z}=0$ and $\Gamma _{\parallel }=0$, 
we have $\sigma _{pz}(\tau )=\sigma _{zp}(\tau )=0,$ and then $S(\tau) = \mathrm{Re} 
[A^{2}\sigma_{pp}(\tau )+B^{2}\sigma _{zz}(\tau )]-\langle I_{1}\rangle ^{2}$. Finally, the 
total
noise spectrum, $S(\omega )=S_{0}+S_{1}(\omega )+S_{2}(\omega )$, is
\begin{subequations}
\begin{equation}
S_{0}=g_{0}V\coth \left( \frac{V}{2T}\right) +\frac{1}{2}C^{+}(\Delta )\left[
1-\left( \frac{2g_{1}\Delta }{C^{+}(\Delta )}\right) ^{2}\right] ,
\end{equation}
\begin{eqnarray}
S_{1}(\omega ) &=&\frac{(\delta I)^{2}\Gamma _{d}\Delta ^{2}}{(\omega
^{2}-\Delta ^{2})^{2}+4\Gamma _{d}^{2}\omega ^{2}},  \label{s1} \\
S_{2}(\omega ) &=&\left[ 1-\left( \frac{2g_{1}\Delta }{C^{+}(\Delta )}
\right) ^{2}\right] \frac{[C^{-}(\Delta )]^{2}\Gamma _{d}}{\omega
^{2}+4\Gamma _{d}^{2}},
\end{eqnarray}
\end{subequations}
with $\delta I=4V\sqrt{g_{0}g_{1}}$. The pedestal shot noise, $S_{0}$, is
identical with the previously found result\cite{Shnirman}. For the frequency-relevant noise 
spectrum in a S qubit, the decoupled qubit
dynamics give rise to two distinct components: $S_{1}(\omega)$ stems from the transverse 
qubit dynamics, $\sigma _{p}$, i.e. \emph{elastic and inelastic} measurement-induced qubit
decoherence, which is similar to the calculation of Ref.~\onlinecite{Korotkov} but augmented 
with our specific predictions for $V$- and $T$-dependent parameters, $\Gamma _{d}$ and 
$\delta I$; while $S_{2}(\omega)$ is generated by the longitudinal qubit dynamics 
$\sigma_{z}$ (qubit relaxation), which is exclusively due to \emph{inelastic} processes in 
measurement. Clearly, our $S_1(\omega)$ is similar to Shnirman's Eq.~(35)\cite{Shnirman} 
except for the reduction factor. Moreover, our $S_2(\omega)$ coincides with $C_3(\omega)$ 
[Eq.~(39)]\cite{Shnirman} under the same limit conditions, and also vanishes for $V<\Delta$ 
at $T=0$. In our model, $S_{2}(\omega)$ is at least one order of magnitude smaller than 
$S_{1}(\omega)$, irrespective of $V$, $T$, and $g_1/g_0$ [inset in Fig.~1(a)], indicating 
that the detector spectrum mainly reflects S qubit decoherence behavior. Noticing that our 
different low-bias behavior of $S_1(\omega)$ as compared to Ref.~\onlinecite{Shnirman} is due 
to the coarse-graining assumption, which requires $V \gtrsim \Delta$ to guarantee accurate 
noise spectrum at $\omega\sim \Delta$.

\begin{figure}[htb]
\includegraphics [width=8.5cm,height=5.5cm,angle=0,clip=on]{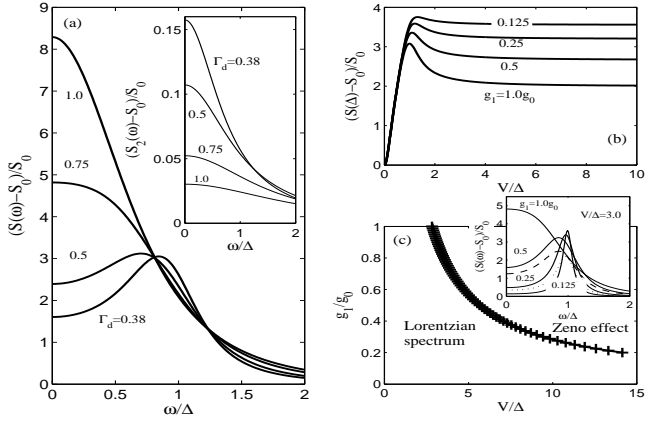}
\caption{(a) $S(\protect\omega)$ as function of frequency $\protect\omega/\Delta$ for various 
$\Gamma_d$ values. The parameters we use in
calculation are: $g_0=g_1=0.25$ and $T=0.1\Delta$; (b) The
peak-to-pedestal ratio vs. bias-voltage; (c) Ranges of various
functional forms for the noise spectrum for a given QPC transparency $g_0=0.25$. Inset: 
$S(\omega)$ vs. $\omega/\Delta$. Solid lines denote results for various $g_1/g_0$ values at 
$g_0=0.25$. The dashed and dotted curves are for $g_1=g_0$ but lower values of $g_0=0.125$ 
and $0.06$, respectively.} \label{fig1}
\end{figure}

Figure 1(a) plots measurement-induced qubit decoherence effects on the noise
spectrum in the maximal meter-response case $g_1=g_0=0.25$. It may be observed that with 
increasing $\Gamma _{d}$ (corresponding voltages $V/\Delta =1.5$, $2.0$, $3.0$, and $4.0$), 
$S(\omega)$ changes from a Lorentzian-type
function, with a peak located at the qubit Rabi frequency $\omega =\Delta$, to the
spectrum shape
centered at $\omega =0,$ indicating the Zeno effect regime. This behavior can be understood 
as follows: (1) a low voltage applied to QPC can not provide sufficient dissipation (small 
$\Gamma_d$) to suppress the Rabi oscillation of a S qubit, thus a peak occurs in the noise 
spectrum contrary to the previous prediction of Ref.~\onlinecite{Shnirman} as mentioned 
above; on the other hand, (2) a high voltage can freeze the qubit due to enhanced relaxation, 
indicating appearance of the Zeno effect. The specific value of voltage dividing the two 
regimes naturally depends on the QPC-qubit coupling for a given QPC transparency $g_0$. 
Based on Eq.~(\ref{s1}), the
transition occurs under the condition $\Gamma _{d}>\Delta /\sqrt{2}$, which leads to a rough
transition boundary involving $\chi /W$ and $V$ ($\log \frac{g_{1}}{g_{0}}\propto -\log V$) 
as shown in Fig.~1(c) for the QPC with a fixed transparency $g_0=0.25$. 
The Lorentzian spectrum regime extends to a wider range of $V$ for systems with weaker 
detector-qubit coupling $\chi$. The inset of Fig.~1(c) exhibits the fact that the spectrum 
changes from the Zeno regime back to the Lorentzian regime as $g_1$ decreases. We should 
notice that the transition boundary depends on $g_0$ due to $g_1\leq g_0$, though $\Gamma_d$ 
is independent of $g_0$. In the high-voltage limit, $V\gg \Delta$, $T$, we can still obtain 
the Lorenzian noise spectrum of detector output via applying a QPC with very low conductance, 
$g_0\ll \Delta/V$, to guarantee $\Gamma_d\ll \Delta$. This remark is shown by dashed and 
dotted lines in the inset of Fig.~1(c).     

Another interesting feature is the quantum upper bound of $4$ for
signal-to-noise ratio (SNR) in the qubit measurement, which was first pointed out by Korotkov 
and Averin
in the high voltage limit\cite{Korotkov}. From Fig.~1(b), we observe that (1) the SNR 
increases rapidly initially with rising voltage, and reaches a maximum value less than $4$ 
(depending on $g_{1}$) around $V=\Delta$; (2) increasing qubit-QPC coupling $g_{1}$ results 
in reduction of the SNR due to enhancement of $S_0$, and the SNR becomes $2$ in the high 
voltage limit in the case of maximal response $g_1=g_0$. These results indicate that low QPC 
transparency $g_0$, weak meter-qubit coupling, $g_{1}$, and $V\geq \Delta$
are necessary for an efficient meter.

The efficiency of a meter also depends decisively on temperature\cite{Korotkov}. The spectrum 
peak at $\omega=\Delta$ in Fig.~1(a) will gradually disappear with increasing $T$ (not shown 
here), which is ascribed to enhanced detector-induced decay, $\Gamma_d$. From 
Eq.~(\ref{dissipation}), $\Gamma_d$ is independent of $V$ at $T< \Delta/2$ and $V<\Delta$, 
implying that the QPC is always a good meter for a S qubit at low temperature and 
bias-voltage.           

\begin{figure}[htb]
\includegraphics [width=8.cm,height=4.cm,angle=0,clip=on]{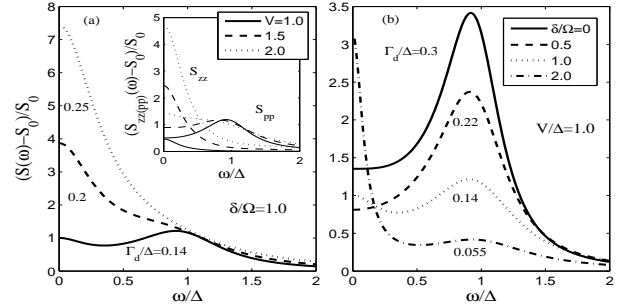}
\caption{(a) $S(\protect\omega)$ vs. $\protect\omega/\Delta$ for an
asymmetric qubit with $\protect\delta=\Omega$ at various
bias-voltages; (b) $S(\protect\omega)$ for different qubits at
$V=1.0 \Delta$. Other parameters: $g_0=g_1=0.25$, $T=0.1
\Delta$.} \label{fig2}
\end{figure}

The noise spectra for asymmetric (A) qubits are summarized in Fig.~2. In
contrast to the S qubit, the A qubit \emph{longitudinal} dynamics start to play
a dominant role in the spectrum around $\omega =0$ when $V\geq \Delta $ [$S_{zz}$ in the
inset of Fig.~2(a)], making $S(\omega )$ approach a zero-frequency maximum more quickly, 
albeit that the transverse spectrum, $S_{pp}$, still has a peak at $\omega =\Delta $. This is 
to say that \emph{breakdown of the resonance condition} causes the A qubit measurement to 
enter into the Zeno effect regime more quickly, even
with $\Gamma _{d}\ll \Delta $. 

In conclusion, we have analyzed the noise output spectrum of a QPC-qubit
measurement system for arbitrary voltage and temperature conditions using a
general quantum-operator Langevin equation approach to derive Bloch equations for the system. 
In contrast to the results in Ref.~\onlinecite{Stace}, which improperly ignores quantum 
interference between {\em elastic} and {\em inelastic} processes of measurement, our 
calculations indicate that qubit oscillations always cause a peak in the QPC
noise spectrum at the Rabi frequency, $\Delta$, for a S qubit, because the qubit coherence
dynamics dominate contributions to the noise spectrum (if the condition $\Gamma _{d}<\Delta /
\sqrt{2}$ is satisfied), which agrees with previous studies in the high voltage 
limit\cite{Korotkov}. However, the coherent peak at the Rabi frequency is suppressed by the 
relaxation
dynamics of an A qubit even when $\Gamma _{d}\ll \Delta$. 
Our analyses provide physical insight into qualitative criteria for design of an efficient 
detector: (1) low transparency $g_0$ of a QPC; (2) weak meter-qubit coupling,
$g_{1}$; (3) small asymmetry ratio, $\delta /\Omega$; (4) relatively low temperature. 

This work was supported by Projects of the National Science Foundation of
China and the Shanghai Municipal Commission of Science and Technology. NJMH
was supported by the DURINT Program administered by the US Army Research Office, DAAD Grant
No.19-01-1-0592.


\begin{thebibliography}{99}

\bibitem{Buks} {E. Buks, R. Schuster, M. Heiblum, D. Mahalu, and V. Umansky,
Nature \textbf{391}, 871 (1998); D. Sprinzak, E. Buks, M. Heiblum, and H. Shtrikman, Phys. 
Rev. Lett. \textbf{84}, 5820 (2000); M.A. Kalish, M. Heiblum, A. Silva, D. Mahalu, and V. 
Umansky, Phys. Rev. Lett. \textbf{92}, 156801 (2004).}

\bibitem{Schoelkopf} {R.J. Schoelkopf, P. Wahlgren, A.A. Kozhevnikov, P.
Delsing, and D.E. Prober, Science \textbf{280}, 1238 (1998).}

\bibitem{Nakamura} {Y. Nakamura, Yu.A. Pashkin, and J.S. Tsai, Nature
\textbf{398}, 786 (1999).}

\bibitem{Mozyrsky} {D. Mozyrsky and I. Martin, Phys. Rev. Lett. \textbf{89},
18301 (2002).}

\bibitem{Buttiker}{S. Pilgram and M. B\"uttiker, Phys. Rev. Lett. {\bf 89}, 200401 (2002); 
A.N. Jordan, M. B\"uttiker, cond-mat/0505044.}

\bibitem{Korotkov} {A.N. Korotkov, Phys. Rev. B \textbf{63}, 85312 (2001);
A.N. Korotkov and D.V. Averin, Phys. Rev. B \textbf{64}, 165310 (2001); R. Ruskov and A.N.
Korotkov, Phys. Rev. B \textbf{67}, 75303 (2003).}

\bibitem{Goan} {H.S. Goan, G.J. Milburn, H.M. Wiseman, and H.B. Sun, Phys.
Rev. B \textbf{63}, 125326 (2001); H.S. Goan and G.J. Milburn, Phys. Rev. B \textbf{64}, 
235307
(2001).}

\bibitem{Shnirman} {A. Shnirman, D. Mozyrsky, and I. Martin,
cond-mat/0211618; Europhys. Lett. \textbf{67}, 840 (2004).}

\bibitem{Bulaevskii} {L.N. Bulaevskii, M. Hru\v ska, and G. Ortiz, Phys.
Rev. B \textbf{68}, 125415 (2003).}

\bibitem{Stace} {T.M. Stace and S.D. Barrett, Phys. Rev. Lett. \textbf{92},
136802 (2004); cond-mat/0309610.}

\bibitem{Li} {X.Q. Li, P. Cui, and Y.J. Yan, Phys. Rev. Lett. \textbf{94},
66803 (2005).}

\bibitem{dispute} {D.V. Averin and A.N. Korotkov, cond-mat/0404549; T.M.
Stace and S.D. Barrett, cond-mat/0406751.}

\bibitem{Dong} {B. Dong, N.J.M. Horing, and H.L. Cui, Phys. Rev. B {\bf 72}, 165326 (2005).}

\bibitem{NFDT} {D. Rogovin and D.J. Scalapino, Ann. Phys. (N.Y.) \textbf{86}, 1 (1974); E.V. 
Sukhorukov, G. Burkard, and D. Loss, Phys. Rev. B \textbf{63}, 125315 (2001).}

\bibitem{QRT} {M.O. Scully and M.S. Zubairy, \emph{Quantum optics},
(Cambridge University Press, Cambridge, 1997).}

\end{thebibliography}
\end{document}